\documentclass[aps, prl, twocolumn, longbibliography]{revtex4-2}
\usepackage{soul}
\usepackage{graphicx}
\usepackage{bm}
\usepackage[normalem]{ulem}
\usepackage{amsbsy,amsmath}

\usepackage{palatino}
\usepackage[utf8]{inputenc}
\usepackage{epsf}
\usepackage{amsmath,amssymb}
\usepackage{calc}

\usepackage{float}

\usepackage[usenames,dvipsnames]{xcolor}

\usepackage{graphicx}
\usepackage{hyperref}

\newcommand{\ben}{\begin{equation}}
\newcommand{\een}{\end{equation}}
\newcommand{\bea}{\begin{eqnarray}}
\newcommand{\eea}{\end{eqnarray}}

\def\ket#1{\vert#1\rangle}

\def\sss{\scriptscriptstyle\rm}

\def\1s{_{1,\sss S}}
\def\2s{_{2,\sss S}}

\def\s{_{\sss S}}
\def\xc{_{\sss XC}}

\def\H{_{\sss H}}

\def\ext{_{\rm ext}}

\def\br{{\bf r}}

\begin{document}

\title{Reformulation of Time-Dependent Density Functional Theory for Non-Perturbative Dynamics:  The Rabi Oscillation Problem Resolved}
\author{Davood B. Dar}
\affiliation{Department of Physics, Rutgers University, Newark 07102, New Jersey USA}
\author{Anna Baranova}
\affiliation{Department of Physics, Rutgers University, Newark 07102, New Jersey USA}
\author{Neepa T. Maitra}
\affiliation{Department of Physics, Rutgers University, Newark 07102, New Jersey USA}
\email{neepa.maitra@rutgers.edu}

\date{\today}

\begin{abstract}
Rabi oscillations have long been thought to be out of reach in simulations using time-dependent density functional theory (TDDFT), a prominent symptom of the failure of the adiabatic approximation for non-perturbative dynamics.  We present a reformulation of TDDFT which requires response quantities only,  thus enabling an adiabatic approximation to predict such dynamics accurately because the functional is evaluated on a density close to the ground-state, instead of on the fully non-perturbative density. Our reformulation applies to {\it any} real-time dynamics, redeeming TDDFT far from equilibrium. Examples of a resonantly-driven local excitation in a model He atom, and charge-transfer in the LiCN molecule are given.
\end{abstract}

\maketitle
While the balance between accuracy and efficiency makes time-dependent density functional theory (TDDFT) one of the most successful methods for predictions of molecular spectra and response~\cite{RG84,TDDFTbook12,Carstenbook,M16,Herbert23}, the difficulty in obtaining functional approximations that perform reliably beyond the response regime has dogged its general use in applications where the system is driven far from equilibrium~\cite{LGIDL20,LM23}. Advances in experiments and in technologies involving non-perturbative electron dynamics, triggered for example by laser fields or collisions with ions, give urgency to solving this problem, especially given the dearth of alternative computationally feasible methods on complex systems. 
In some situations, the TDDFT simulations nevertheless give useful mechanistic information and sometimes give results that qualitatively match the experiment~\cite{RFSR13,XCZCY24,UAC18,S21}
but in others, such as scattering~\cite{GWWZ14,QSAC17,SLWM17} and pump-probe spectroscopy~\cite{HTPI14, GBCWR13,RN12,RN12c},  TDDFT can give large errors,  and even completely fail, such as for Rabi oscillations~
\cite{RB09,HTPI14,FHTR11,EFRM12,LFSEM14,FLSM15} or long-range charge-transfer dynamics~\cite{FERM13,M17,RN11,LM21}. 

In the non-perturbative regime, TDDFT operates via the time-dependent Kohn-Sham (TDKS) equations, in which the many-body effects are mapped to a one-body potential, the exchange-correlation (xc) potential. 
The root of the failures is the adiabatic approximation to the xc functional~\cite{LM23}, which is unable to capture step and peak features that are a signature of memory-dependence of the exact xc functional~\cite{DLM22}. 
While the exact $v\xc[n; \Psi(0),\Phi(0)](\br,t)$ depends on the history of the density $n(\br, t'<t)$, the initial interacting state $\Psi(0)$ and the choice of the initial KS state $\Phi(0)$,  this dependence is neglected in adiabatic approximations that insert the instantaneous density $n(\br,t)$ into a ground-state approximation $v\xc^{\rm g.s.}[n(t)](\br)$. 
The non-adiabatic features play a critical role in correcting spurious frequency-shifts in spectral peaks of systems driven out of a ground-state~\cite{FLSM15,HTPI14,RN12,RN12c} that occur in simulations using an adiabatic approximation. These shifts have an especially grave consequence for resonantly-driven systems, causing the adiabatic TDKS simulation to detune itself from the driving frequency.  Even an adiabatically-exact approximation, meaning one where the exact ground-state functional is used in the TDKS propagation, fails~\cite{FM14,LM23}. It is in fact surprising that there are situations where adiabatic TDKS predictions are qualitatively reasonable { in the non-perturbative regime}, given that the xc functional approximation is being evaluated on a fully non-equilibrium density where the underlying true and KS wavefunctions are typically very far from any ground state.

 Finding a practical non-adiabatic approximation has proven elusive~\cite{LM23}. Developing improved functionals for excitations in the linear response regime has so far been more successful, e.g. 
{incorporating exact-exchange to improved Rydberg spectra and charge-transfer excitations~\cite{TH00,SKB09, BLS10,K17b,M17}, including frequency-dependence to yield double-excitation frequencies and oscillator strengths ~\cite{MZCB04,M22,DM23},  
including long-ranged kernels to capture excitonic spectra~\cite{RORO02,SOR03,MSR03,CBR20,SLKSU21}, and current-density dependence for relaxation and dissipation from electron viscosity~\cite{VK96,VUC97}}. However, the search for practical memory-dependent functionals that contain the requisite non-adiabatic features for non-perturbative dynamics has so far come up dry.

Here, we present a reformulation of TDDFT that applies to non-perturbative electron dynamics while
requiring xc functionals only in the linear and quadratic response regimes. This means that, instead of having to evaluate the xc functionals on the fully non-equilibrium system,  they are evaluated always close to the ground-state, and thus are far more amenable to adiabatic approximations. The same adiabatic functional performs far better in this response-reformulated TDDFT (RR-TDDFT) than it does in the traditional TDKS scheme. A special case of this approach was shown in earlier work on Ehrenfest dynamics~\cite{LM21},  and is related to how electron-nuclear dynamics with TDDFT is often implemented,
but here we show the approach can be extended to general non-perturbative problems, resolving the problem of missing Rabi oscillations and long-range charge-transfer dynamics in the TDKS approach,  as we demonstrate with two examples.

The theorems of TDDFT~\cite{RG84} tell us that from the one-body density $n(\br,t)$ one can extract all observables for a system evolving in the time-dependent many-body Schr\"odinger equation
\ben
i \partial_t \ket{\Psi} = (H^{(0)} + V^{\rm app}(t) )\ket{\Psi}; 
\label{eq:TDSE}
\een
where $H^{(0)} = T + W + V\ext^{(0)}$ is the sum of the kinetic energy operator, electron-electron interaction, and static external potential due to the nuclei, respectively, and $V^{\rm app}(t) = \int d^3r v^{\rm app}(\br,t)\hat{n}(\br)$ is a one-body local (multiplicative) potential operator representing an externally applied field, with $\hat{n}(\br)$  the one-body density-operator.
{We use atomic units ($e^2 = \hbar = m_e = 1$) unless otherwise stated.}
In the standard TDDFT procedure, the system is mapped to the non-interacting KS system that reproduces the exact interacting density $n(\br,t)$ with a set of orbitals that evolve under the TDKS equations:
\ben
(-\nabla^2/2 + v\s(\br,t))\phi_i(\br,t) = i \partial_t\phi_i(\br,t),
\label{eq:TDKS}
\een
where $v\s(\br,t) = v\ext^{(0)}(\br,t) + v^{\rm app}(\br,t) + v\H(\br,t) + v\xc(\br,t)$. Here $v\H(\br,t)$ is the Hartree potential, a functional of the instantaneous density, while $v\xc(\br,t) = v\xc[n; \Psi(0),\Phi(0)](\br,t)$ has the memory-dependence whose neglect in usual approximations leads to errors and failures, as discussed earlier.

Instead, RR-TDDFT bypasses the solution of the TDKS orbitals and solves for a set of TD expansion coefficients of the many-body state, but without needing to actually find the state.  We expand the time-dependent physical many-body wavefunction in terms of the (unknown) many-body eigenstates: $\ket{\Psi(t)} =\sum_n C_n(t)\ket{\Psi_n}$, where $\ket{\Psi_n}$ satisfies the static many-body equation: $H^{(0)}\ket{\Psi_n} = E_n\ket{\Psi_n}$. Inserting this into Eq.~(\ref{eq:TDSE}) gives

\ben
i  \dot{C}_m(t) = E_m C_m(t) + \sum_n V^{\rm app}_{mn}(t) C_n(t)
\label{eq:Cdot}
\een
where the sum goes over all the eigenstates and
\ben
V^{\rm app}_{mn}(t) = \langle \Psi_m\vert V^{\rm app}(t)\vert \Psi_n\rangle = \int d^3 r v^{\rm app}(\br,t)\rho_{mn}(\br)
\label{eq:vapp}
\een
with $\rho_{mn}(\br) = N\int d^3r_2..d^3r_N \Psi^*_m(\br,\br_2..\br_N)\Psi_n(\br,\br_2..\br_N)$ being the transition-density and $N$ the number of electrons.
The time-dependent one-body density can be extracted from 
\ben
n(\br,t) = \sum_{n,m}C_n^*(t)C_m(t)\rho_{nm}(\br)
\label{eq:density}
\een
We now argue that Eqs.~\ref{eq:Cdot}--\ref{eq:density} provide a route to obtaining all observables of a non-perturbative real-time dynamics from just TDDFT response properties. First, invoking the Runge-Gross theorem, all observables can be obtained from the initial interacting state $\ket{\Psi(0)}$ and the time-evolving density $n(\br,t)$.  Eq.~(\ref{eq:density}) provides $n(\br,t)$, which requires solution of the coupled time-evolution equations, Eqs.~(\ref{eq:Cdot}) for the coefficients. To solve these, we need:
\newline
(i) the initial coefficients $C_m(0)$ which are obtained from expanding the initial many-body state $\ket{\Psi(0)}$ in terms of the many-body eigenstates of $H^{(0)}$. It is important to note that the interacting eigenstates themselves are {\it not} required, only knowledge of which states are occupied and with what amplitudes, which would be determined by the physics of the initial conditions of the problem. Often this is just the ground-state, in which case $C_0(0) = 1, C_{m\neq 0}(0) = 0$. 
\newline
(ii) energies $E_m = E_0 + \omega_m$ that can be obtained from adding frequencies from TDDFT linear response~\cite{PGG96,C95} $\omega_m$ to the ground-state DFT energy $E_0$, and
\newline
(iii) the transition-density $\rho_{mn}(\br)$ which can be obtained from linear response TDDFT~\cite{PGG96,C95,TCR09} for ground-excited transitions, and quadratic response for excited-excited transitions~\cite{PRF13}.

\begin{figure}[h]
    \centering
    \includegraphics[width=\linewidth]{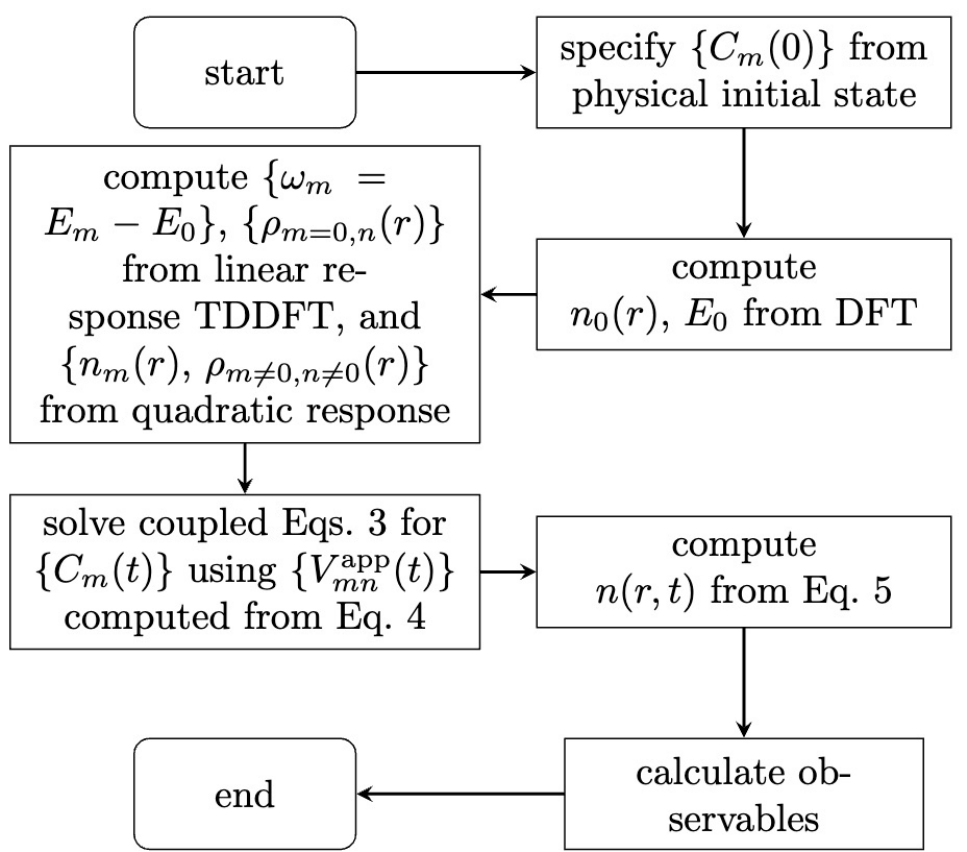}
    \caption{ Flowchart of the RR-TDDFT procedure }
    \label{Flowchart}
\end{figure}

With the ingredients in (ii)--(iii) all obtained from ground-state DFT, linear and quadratic TDDFT response, the time-dependent density Eq.~(\ref{eq:density}) can be obtained,   and hence all observables~\cite{RG84}. 
 We again stress that neither the time-evolving many-body wavefunction, nor the many-body eigenstates are needed in this procedure. 
The idea may be seen as similar in spirit to the time-dependent configuration method, but here linear and quadratic response TDDFT is used to obtain the static electronic structure quantities and we never actually find the wavefunction. 
The procedure is illustrated in Fig~\ref{Flowchart}.

A common situation is when the applied field is a laser field modelled by a potential $v^{\rm app}(\br,t) = {\boldsymbol{\mathcal E}}(t)\cdot \br$, and the observable of interest is the dipole moment ${\bf d}(t)$. In this case,  Eqs~(\ref{eq:Cdot})--(\ref{eq:density}) simplify to
\bea
\nonumber
i \dot C_m(t) &=& E_m C_m(t) + {\boldsymbol{\mathcal E}}\cdot\sum_n {\bf d}_{mn} C_n(t)\\
{\bf d}(t) &=& \sum_n\sum_m C_n^*(t)C_m(t) \,{\bf d}_{mn}
\label{eq:simpler}
\eea
where ${\bf d}_{m=n}$ are the dipole moments of the excited state $\Psi_n$, available from response TDDFT~\cite{F01,FA02}, ${\bf d}_{m\neq n}$ are the transition dipoles between excited states, available from linear response when one of the state is the ground-state, and otherwise from quadratic response~\cite{PGG96,C95,TCR09,PRF13}.  

Comparing with the standard TDDFT procedure based on the TDKS equations, Eq.~\ref{eq:TDKS},
the saliant advantage of RR-TDDFT is
that the domain on which the adiabatic xc functional approximations are evaluated (ground states and their linear and quadratic responses), is far closer to the domain in which { the approximation was} derived. In contrast, the xc functional in TDKS with an adiabatic approximation, $v\xc^{A}[n; \Psi(0),\Phi(0)](\br,t) = v\xc^{\rm g.s.}[n(t)](\br)$, applies a ground-state functional in a domain on the left-hand-side which evolves far from any ground-state. Such an approximation is unlikely to be accurate. On the other hand, in RR-TDDFT, there are three xc objects: $v\xc^{\rm g.s.}[n_{\rm g.s.}](\br)$ (needed for $E_0$ and the KS orbitals and excitation energies that the linear response builds upon), $f\xc[n_{\rm g.s.}](\br,\br',\omega)$ (the central xc kernel in linear response TDDFT), and $g\xc[n_{\rm g.s.}](\br,\br',\br'',\omega,\omega')$ (the second-order response kernel). The latter two are related to functional derivatives of $v\xc[n,\Psi(0),\Phi(0)](\br)$  evaluated on a ground-state density, so the domain involves only small perturbations around a ground-state density.

Another fundamental difference is in the role of the initial state. Similar to TD wavefunction methods, the physical interacting initial state is a key input in RR-TDDFT, but in the TDKS approach, it appears only { implicitly} through the functional dependence of $v\xc[n; \Psi(0), \Phi(0)](\br,t)$.  { This initial-state dependence unknown, and is, in practise,} neglected in TDKS, since adiabatic functionals depend only on the instantaneous density. {However,} the exact xc functional varies significantly when the system starts in different initial states even if they all have the same one-body density~\cite{EM12,FNRM16}.  In a sense, the initial interacting state  plays a more prominent, and conceptually easier, role in RR-TDDFT than in TDKS since it appears directly as an initial condition in the evolution equations, rather than in an unknown functional-dependence; { more precisely, its expansion coefficients in terms of the unknown many-body eigenstates appear, the spatial dependence of the initial wavefunction and the many-body eigenstates themselves are not needed.}
 The initial condition in TDKS is, instead,  the KS initial state $\Phi_0$.  Different choices of $\Phi_0$ give different xc potentials, and the adiabatic approximation gives significantly varying errors for different choices~\cite{LM20b,EM12,SLWM17}. Adiabatic approximations perform particularly poorly when the rank of the true interacting density-matrix evolves significantly (e.g. $\Psi(t)$ going from close to a single Slater determinant to a singly-excited singlet state), because the TDKS state $\Phi(t)$ cannot change its rank. 
These challenging considerations are moot in RR-TDDFT.

Effectively, RR-TDDFT separates out the time and space-dependence of observables, and  in doing so reduces the complexity in the xc effects arising from the inherent entanglement of time- and spatial-non-locality~\cite{DBG97, Vignalechap}. 
This leads to quite different numerical considerations in the two approaches:  the RR-TDDFT trades a self-consistent solution of the set of $N$ partial-differential TDKS equations in space and time where $N$ is the number of electrons, for a self-consistent solution of a set of $M$ ordinary differential equations in time, where $M$ represents the anticipated number of many-body states that will likely be occupied during the dynamics. RR-TDDFT also needs to solve linear response equations for $M$ energies, densities, and $M(M-1)$ couplings, some of which require quadratic response.  
{ We can estimate $M$ from considering properties of the applied field, such as intensity, frequencies, and polarizations. Ultimately for a general non-perturbative situation, we may need to do a convergence study where more states are added until there is no significant change in the resulting observables}. In scenarios where a very large number of states are likely to be involved, RR-TDDFT may become unfeasible.
But in many scenarios { $M$ may be in the single-digits (e.g. just 2 for resonant driving), regardless of the number $N$ of electrons,  and RR-TDDFT, in addition to its much more reliable predictions, {will} offer a computational advantage over TDKS due to its ordinary rather than partial differential equation nature.
For computing excited-to-excited state couplings, we note that quadratic response may be circumvented by approximating these from linear response akin to the auxiliary wavefunction method employed to compute derivative couplings between excited states\cite{TCR09,OBFS15}. 

We now  give two examples to demonstrate how adiabatic functionals achieve Rabi oscillations when used within RR-TDDFT, while completely failing within TDKS.  

Our first example is a one-dimensional { (1D)} Helium atom (1D He) with soft-Coulomb interactions, studied before in this context~\cite{RB09,FHTR11,EFRM12, LFSEM14}: $v\ext^{(0)} = -2/\sqrt{1 + x^2}$, contained in a box of size -40 a.u. to 40 a.u.
{ The real-space octopus code~\cite{octopus} was used in this example.}
We apply a field $\mathcal{E}(t) = 0.00667\sin(\omega t)$ to the ground-state, where $\omega$ is resonant with the first singlet excitation, $\omega = \omega^{ex} = 0.5336$a.u. With the transition dipole moment of $\mu_{01} =1.106$a.u., this gives a Rabi period of $T_R$ where $T_R/2 = \frac{\pi}{0.00667\mu_{01}} = 425.9$a.u. 
The observable we are interested in is the dipole moment, $d(t)=\int x n(x,t) dx$ where $n(x,t)=\sum_{i occ} \vert\phi_i(x,t)\vert^2$ is obtained from the solution of Eq.~(\ref{eq:TDKS}) for TDKS and  from Eqs.~(\ref{eq:simpler})
for RR-TDDFT. Owing to the spatial symmetry of the ground and first excited states resulting in a zero permanent dipole moment, the dipole moment from Eq.~(\ref{eq:density}) {(or Eq.~\ref{eq:simpler})} simplifies to $d(t)=2\Re\left[C^*_0(t)C_1(t)\right]\mu_{0,1}$. 

The top left panel of Fig.~\ref{fig:1d_dip} depicts the exact dipole moment dynamics obtained the exact solution from the TDSE, with the expected Rabi oscillation. The second and third plots in the top panels show the result of TDKS evolution with the exact exchange (EXX) approximation  and two different driving frequencies: the driving frequency in the second plot is as in the exact dynamics, while in the third, the frequency is instead that of the
 excited state predicted by EXX linear response, $\omega^{\mathrm{EXX}} = 0.5488$ a.u. 
As has been observed in earlier work~\cite{FHTR11}, in both cases a significant deviation from the exact behavior is evident, underscoring the failure of TDKS to accurately replicate the observed dynamics. The dipole envelopes falsely suggest a Rabi-like oscillation albeit at wrong frequencies (the expected half-Rabi period calculated from   $\mu_{01}^{\rm EXX}=-1.0924$ a.u. gives $ T_R^{\rm EXX}/2 = 431.2$ a.u.): the density at the minimum of the dipole moment is not that of the EXX excited state, { as shown in} the bottom right panel, to be discussed more shortly. 

Turning now to our RR-TDDFT method, the first plot in the bottom panel  shows the result of applying the pulse used in the exact case in Eqs.~\ref{eq:simpler} , using the energies and transition densities given by EXX. We observe the expected detuned Rabi oscillation, and only a partial population transfer, due to the mismatch of $\omega^{\rm EXX}$ with the driving $\omega$. Applying the field instead at $\omega^{\rm EXX}$, displays  a true Rabi oscillation as shown in the middle plot. There is a full population inversion at $T_R^{\rm EXX}/2$, and the densities in the right-most plot verifies this. The density at this time has the same shape as the excited state density, unlike that of the TDKS density at what looks like its half-Rabi time. This plot also shows the excited-state density $n_{\rm EXX}^*(x)$ computed from a response calculation with EXX~\cite{F01,FA02,BM24}, which is very close to the exact excited state density $n^*(x)$. Thus, while EXX failed to produce a Rabi oscillation when used within TDKS, this same functional approximation succeeded when used within RR-TDDFT.

\begin{figure}
    \centering
    \includegraphics[width=0.5\textwidth]{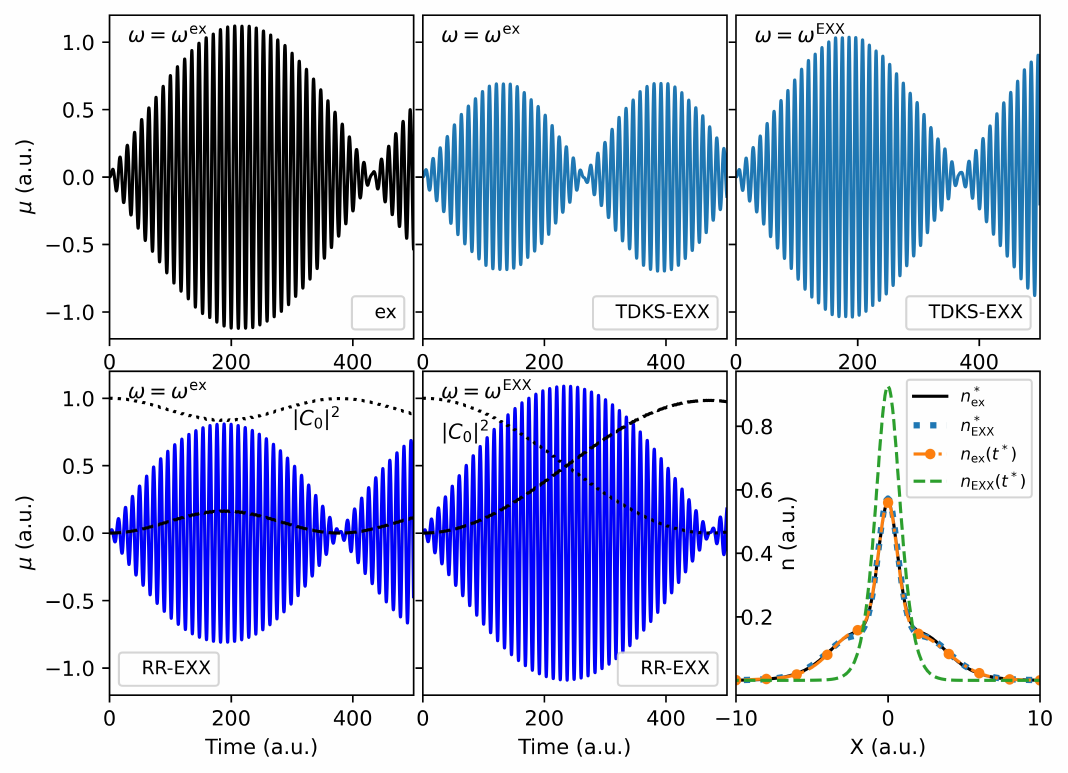}
    \caption{Resonantly-driven dipole dynamics  in 1D He. Top panels: (Left) Numerically exact result calculated using TDSE  with a pulse, $\mathcal{E}(t) = 0.00667\sin(\omega t)$ and $\omega=\omega^{\rm ex} = 0.5336$a.u.; (Middle) TDKS with EXX functional;  (Right) TDKS-EXX driven by $\omega^{\rm EXX} = 0.5488$a.u.  Bottom panels: (Left and Middle) Dipole moments calculated from RR-TDDFT with EXX driven at $\omega^{\rm ex}$ and $\omega^{\rm EXX}$ respectively. Also shown in dotted and dashed lines are the populations of the ground and the first singlet excited states, $\vert C_0(t)\vert^2$ and $\vert C_1(t)\vert^2$ (unlabelled).   (Right) The exact $n^*_{\text{ex}}(x)$ (black solid) and EXX $n_{\text{EXX}}(x)$ (blue dotted) excited-state densities calculated from static (response) calculations,  compared with the exact time-evolved densities $n^*_{\text{ex}}(x,t^* = 425.9{\rm a.u.})$ (orange with circle) and EXX, $n_{\text{EXX}}(x,t^* = 364.35{\rm a.u.})$ (green dashed). }
    \label{fig:1d_dip}
\end{figure}

We now turn to the case of the lithium cyanide (LiCN) molecule, which has been studied in the past as a test system for light-driven dipole switching~\cite{KKS05,RN11}: the degenerate second (S$_2$) and third (S$_3$) excited states have a dipole moment along the bond-axis ($\hat{z}$) opposite to that in the ground-state. Applying a laser pulse resonant with the excitation frequency along the $\hat{x}$ ($\hat{y}$) direction, which coincides with the direction of the transition dipole to the  S$_2$ (S$_3$) states respectively, drives the transition to the S$_2$ (S$_3$) state with a concomitant large change in the $z$-dipole moment. 
The known failure of the TDKS simulation to accurately describe this dipole switching is one of the prime examples of limitations of adiabatic approximation~\cite{RN11,RN12,FERM13,LGIDL20,LM23}. Here we show that the same adiabatic approximations perform well when applied instead within the RR-TDDFT approach. We use the NWChem~\cite{nwchem} code to perform the real-time TDKS calculations, and its linear response module used to extract the ingredients in Eqs.~\ref{eq:Cdot}--\ref{eq:density} for the RR-TDDFT, truncated to two states: ground S$_0$ and excited S$_2$-state energies, dipole moments, and the transition dipole moment between these states.

We take a short enough pulse that the nuclei may be treated statically during the evolution, fixed at their equilibrium geometry, $R_{\rm Li–C} = 3.683$ a.u. and $R_{\rm C–N} = 2.168$ a.u~\cite{KKS05}. The applied field is a resonant $\pi$-pulse~\cite{HJ94} along the $\hat{x}$-direction, 
\begin{equation}
v^{\rm app}(\br,t) = x f_{0}\sin^2 \left(\frac{\pi t}{2\sigma}\right) \sin(\omega t)
\label{eq:LiCNfield}
\end{equation}
where $\omega$ is the excitation frequency of the S$_2$ state, $\sigma$ is the half-width of the pulse envelope and the amplitude $f_{0}=\frac{\pi}{\sigma \vert \mu_{0,2;x}\vert}$ 
with $\mu_{0,2;x}$, the $x$-transition dipole moment between states $S_0$ and $S_2$. (For a two-state problem, this pulse achieves population inversion by time $T= 50$fs).
We will take the reference (``exact") calculation as the time-dependent CISD(10,15)/6-31G* simulation of Fig. 3 in Ref.~\cite{RN11}, which applied this $\pi$-pulse at resonant frequency $\omega^{\rm ex} = 6.8$eV, close to the linear response CISD value of 6.77 eV, and $\sigma = 25$ fs, such that a full population inversion is achieved at around 38 fs.

The top panel of Fig.~\ref{fig:LiCN} shows the $z$-dipole moment  $\mu_z$ when driven by the $\pi$-pulse of Eq.~\ref{eq:LiCNfield}, as predicted from TDKS and our RR-TDDFT, using adiabatic PBE~\cite{PBE96} and a tuned BNL (tBNL)~\cite{BN05,LB07} functional, both using the same 6-31G* basis set as the reference CISD. 
The resonant frequencies predicted by linear response with these functionals are  $\omega^{\rm PBE} = 4.31$eV, and $\omega^{\rm tBNL} = 6.80$eV where we tuned the range-separation parameter $ \gamma_{BNL}=0.8$ in order to align the excitation energy of S$_2$ state with the applied frequency.
The complete failure of {both} TDKS simulations is evident in the figure, similar to what was observed in the earlier work~\cite{RN11}~\footnote{Ref.~\cite{RN11} used a different pulse for the TDKS simulations which gave more oscillatory behavior}, and in model system analogs of the problem~\cite{FERM13,FM14,M17} (which used a flat envelope rather than a $\pi$-pulse). In particular, despite the close agreement of the  tBNL linear response frequency with the reference, the real-time TDKS calculation of the dynamics is miserable. In contrast, RR-TDDFT with this functional (RR-tBNL in the figure) is excellent.  RR-TDDFT using PBE still fails, with even less of a response than TDKS-PBE, as shown in the inset. 
 This is because the PBE frequency is severely underestimated due to the charge-transfer nature of the excitation, and with such a weak field, off-resonant to any system frequency, the system is barely disturbed. 

Instead,   RR-TDDFT with PBE achieves dipole switching for a pulse that is resonant with the PBE frequency. 
This is shown in the lower panel of Fig. ~\ref{fig:LiCN} where we again show TDKS and RR-TDDFT with PBE and tBNL functionals, but with the frequency and transition dipole that enter into the resonant pulse Eq.~\ref{eq:LiCNfield} obtained from the corresponding underlying electronic structure. If we did not have a reference calculation, and were relying on the PBE (or tBNL) functional for our description of the system, these would be the $\pi$-pulse parameters we would use to achieve the resonant charge-transfer. 
Now we see that, due to functionals being evaluated on a domain closer to the ground-state, RR-TDDFT with either the PBE or tBNL functional reproduce the dipole-switching well, while these same functionals used in the real-time TDKS scheme fail. 
\begin{figure}
    \centering
    \includegraphics[width=0.45\textwidth]{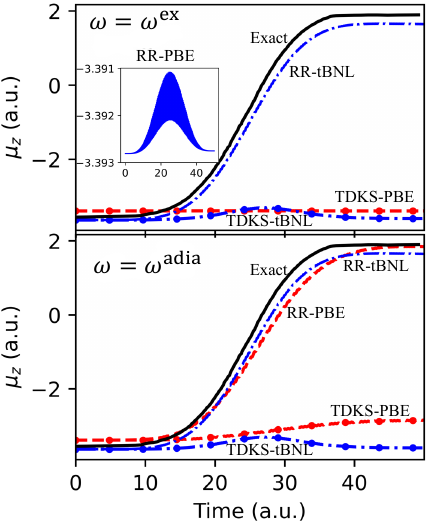}
    \caption{Resonantly-driven charge-transfer in the LiCN molecule:
Top panel: Dipole moment computed from TDKS, and RR-TDDFT, using PBE and tBNL functionals within the 6-31G* basis set. The applied $\pi-$ pulse has the parameters $\omega^{\rm app} = \omega^{\text{ex}}=6.8 eV$, $\sigma=25$ fs and $f_{0}= 0.01019$ and the results are compared with the reference TD-CISD from Ref.~\cite{RN11}. Lower panel: Same quantities as above panel in which the pulse parameters are determined by the corresponding electronic structure. }
    \label{fig:LiCN}
\end{figure}

In summary, our reformulation of real-time TDDFT in terms of response quantities significantly improves electron dynamics far from the ground state when using standard functionals. While these functionals fail to produce Rabi oscillations in the TDKS scheme, they succeed in the RR-TDDFT framework. This is because, unlike in TDKS, the xc functionals in RR-TDDFT are only required in the response regime, where the xc potential is evaluated on densities close to the ground state, aligning with the domain for which the functional approximations were derived.

RR-TDDFT allows us to compute non-perturbative electron dynamics with as much confidence as TDDFT is used in the response regime.
We note that, like TDKS, RR-TDDFT is exact in principle, and need not be limited to using adiabatic approximations.  RR-TDDFT with an adiabatic approximation will work poorly in cases where these approximations are known to fail in linear~\cite{MZCB04,DM23} as well as quadratic~\cite{PRF16,DRM23} response regimes, but given that it has so far proven to be clearer to identify such cases, and to develop improved non-adiabatic response functionals, RR-TDDFT promises to overcome the reliability challenges of TDDFT in the non-perturbative regime.  { Finally, we note that the RR-TDDFT framework could in fact be used in conjunction with any method that provides excited state energies, densities, and transition densities. }

\acknowledgments{Financial support from the National Science Foundation Award CHE-2154829 (AB), the Department of
Energy, Office of Basic Energy Sciences, Division of
Chemical Sciences, Geosciences and Biosciences under
Award No. DE‐SC0024496 (NTM), and the Rutgers Dean’s Dissertation Fellowship (DBD) are gratefully acknowledged. }
\bibliography{main.bib}
\end{document}